\documentclass{article}
\usepackage{graphicx} % Required for inserting images
\usepackage{amsmath, amssymb}

\usepackage{subcaption}
\usepackage{caption}

\usepackage{geometry}
\usepackage{multirow}
\usepackage{hyperref}

\usepackage{threeparttable}

\newcommand\Tstrut{\rule{0pt}{2.6ex}}         % = `top' strut
\newcommand\Bstrut{\rule[-1.3ex]{0pt}{0pt}}   % = `bottom' strut

\usepackage{xcolor}   % safer for colors
\usepackage[normalem]{ulem} % for strikethrough

\newcounter{comments}
\title{A Geometric Graph-Based Deep Learning Model for Drug-Target Affinity Prediction}
\author{Md Masud Rana$^{1}$\thanks{Corresponding author(s): mrana10@kennesaw.edu; ducnguyen@utk.edu}\,, Farjana Tasnim Mukta$^1$, and Duc Duy Nguyen$^{2}\footnotemark[1]$\\
$^1$ Department of Mathematics,
Kennesaw State University, Kennesaw, GA 30144, USA\\
$^2$ Department of Mathematics,
University of Tennessee, Knoxville, TN 37996, USA
}
 \date{}

\begin{document}

\maketitle
\begin{abstract}
    In structure-based drug design, accurately estimating the binding affinity between a candidate ligand and its protein receptor is a central challenge. Recent advances in artificial intelligence, particularly deep learning, have demonstrated superior performance over traditional empirical and physics-based methods for this task, enabled by the growing availability of structural and experimental affinity data. In this work, we introduce DeepGGL, a deep convolutional neural network that integrates residual connections and an attention mechanism within a geometric graph learning framework. By leveraging multiscale weighted colored bipartite subgraphs, DeepGGL effectively captures fine-grained atom-level interactions in protein–ligand complexes across multiple scales. We benchmarked DeepGGL against established models on CASF-2013 and CASF-2016, where it achieved state-of-the-art performance with significant improvements across diverse evaluation metrics. To further assess robustness and generalization, we tested the model on the CSAR-NRC-HiQ dataset and the PDBbind v2019 holdout set. DeepGGL consistently maintained high predictive accuracy, highlighting its adaptability and reliability for binding affinity prediction in structure-based drug discovery.
    
\end{abstract}


\section{Introduction}

Understanding the strength and specificity of protein–ligand interactions is a central challenge in molecular biology and drug discovery. Binding affinity—the quantitative measure of how tightly a ligand binds to its target protein—plays a crucial role in determining the efficacy and selectivity of therapeutic compounds. The binding affinity between a ligand and its protein receptor is often expressed as inhibition constant (Ki), dissociation constant (Kd), or half-maximal inhibitory concentration (IC50). Traditional experimental techniques such as isothermal titration calorimetry (ITC) and surface plasmon resonance (SPR) remain the gold standard for measuring binding affinity but are resource-intensive, time-consuming, and not scalable for high-throughput applications \cite{velazquez2006isothermal, maynard2009surface}. This has fueled a growing demand for efficient computational approaches capable of delivering accurate affinity predictions to accelerate the drug discovery pipeline.

Computational scoring functions have long served as the cornerstone of structure-based drug discovery. These scoring functions are generally classified into physics-based, empirical, and knowledge-based categories. Physics-based methods rely on force-field calculations to estimate binding free energy by accounting for electrostatic and van der Waals interactions \cite{gilson2007calculation, huang2006physics}, while empirical scoring functions are trained on experimental data to weight specific interaction terms \cite{eldridge1997empirical, wang2002further}. Knowledge-based scoring functions derive statistical potentials from structural databases to estimate binding propensities \cite{velec2005drugscorecsd, huang2010inclusion}. In recent years, machine learning (ML) and deep learning (DL) approaches have emerged as powerful alternatives that can learn complex non-linear relationships from large datasets without relying on hand-crafted features \cite{ballester2010machine, durrant2010nnscore}.

ML- and DL-based scoring functions can be broadly divided into sequence-based and structure-based approaches. Sequence-based models such as DeepDTA \cite{ozturk2018deepdta} and DeepDTAF \cite{wang2021deepdtaf} use protein sequences and ligand SMILES as input, enabling high-throughput predictions even in the absence of 3D structural data. However, they often lack spatial resolution, limiting their ability to capture the true nature of molecular interactions. In contrast, structure-based models such as RF-Score \cite{ballester2010machine, li2015improving}, Pafnucy \cite{stepniewska2018development}, and K$_\text{DEEP}$ \cite{jimenez2018k} operate directly on the 3D atomic coordinates of protein–ligand complexes, enabling a more detailed modeling of spatial interactions. These models have demonstrated superior predictive power but also face challenges such as generalization to novel chemical spaces and sensitivity to structural noise.

Graph theory provides a versatile and expressive framework for modeling complex molecular systems, particularly in the context of protein–ligand interactions. Molecules can be naturally represented as graphs, where atoms serve as nodes and bonds or intermolecular interactions correspond to edges. This abstraction enables the encoding of both structural and interaction-level information in a compact form. Graph theory encompasses several branches, including geometric, algebraic, and topological graph theory. Geometric graph theory emphasizes spatial relationships between nodes, capturing the geometric connectivity critical for understanding local atomic environments \cite{bramer2018multiscale, nguyen2017rigidity}. Algebraic graph theory analyzes the spectral properties of graphs using matrices such as the adjacency and Laplacian matrices to extract features related to network connectivity and interaction strength \cite{nguyen2019agl,chen2021algebraic, mukta2025algebraic}. Topological graph theory investigates the embedding of graphs into topological spaces, offering tools such as persistent homology to explore the shape and continuity of molecular structures \cite{wang2020persistent,meng2021persistent}. Recently, geometric graph learning has emerged as a powerful technique for modeling biomolecular complexes by combining graph-based representations with machine learning \cite{jiang2021ggl,rana2023geometric, rana2023gglPPI}. This approach leverages geometric multiscale weighted colored subgraphs (MWCG) that encode local interactions between ligand atoms and protein residues, enabling models to capture both spatial organization and physicochemical context. The MWCG framework assigns edge weights based on Euclidean distances and labels interactions using radial basis functions, offering rotation- and translation-invariant representations for protein flexibility analysis and protein-ligand interactions \cite{bramer2018multiscale, rana2023geometric}. These graph-based methods facilitate the extraction of interpretable and robust features that reflect critical biophysical interactions, thereby enhancing the accuracy and generalizability of binding affinity prediction models.

While recent structure-based deep learning models for protein–ligand binding prediction have shown encouraging results \cite{zheng2019onionnet, wang2021onionnet, seo2021binding, xia2023leveraging}, they often rely on limited feature representations and may not fully capture the geometric and physical principles underlying molecular interactions.
Developing new deep learning frameworks offers the opportunity to address these limitations by incorporating richer structural encodings, integrating physical priors, and improving the interpretability and generalization of the models. 
In this study, we introduce DeepGGL, a deep geometric graph learning model for protein–ligand binding affinity prediction. DeepGGL constructs multi-range weighted bipartite subgraphs that encode chemical and geometric interactions between protein and ligand atoms of specific types. These subgraphs capture interaction patterns across multiple spatial scales, reflecting both local binding determinants and long-range structural context. The descriptors are fed into a deep 2D convolutional neural network with residual connections and attention mechanisms, enabling efficient learning of discriminative interaction features. Unlike prior models that rely solely on local contact maps or voxel grids, DeepGGL integrates multi-scale geometric information to form a holistic view of molecular interactions.
We validate DeepGGL on standard benchmark datasets, including CASF-2013 and CASF-2016, as well as independent test sets such as CSAR-NRC-HiQ and the PDBbind v2019 holdout set. Across all evaluations, DeepGGL demonstrates superior performance compared to state-of-the-art methods. Ablation studies further show the importance of the residual blocks and attention layers in capturing complex interaction patterns and improving generalization. Overall, DeepGGL offers a robust, interpretable, and scalable approach to binding affinity prediction, offering a promising direction for next-generation structure-based scoring functions in drug discovery pipelines.

%%==============================================================%%
%%==============================================================%%
\section{Materials and Methods}

%%==============================================================%%
%%==============================================================%%
\subsection{Datasets preparation}
The primary dataset used for model training in this study is the PDBbind v2016 dataset (\url{http://www.pdbbind.org.cn}), which comprises three overlapping subsets: the general set, the refined set, and the core set, with the latter being a subset of the former. The general set includes all available protein-ligand complexes, while the refined set consists of high-quality complexes with reliable structural and binding affinity data. Each complex in the PDBbind database is annotated with experimentally measured binding affinities, standardized into negative logarithmic scale (pKa) values to ensure consistency across different assay types.

For model evaluation, we employed the widely used CASF-2016 \cite{su2018comparative} and CASF-2013 \cite{li2014comparative} benchmark test sets. The CASF-2016 set, corresponding to the core set of PDBbind v2016, contains 285 high-quality protein-ligand complexes with experimentally validated binding affinities. The CASF-2013 set, derived from the core set of PDBbind v2013, comprises 195 complexes organized into 65 clusters, with binding affinities spanning nearly ten orders of magnitude.

To avoid data leakage between training and evaluation, all complexes included in CASF-2016 and CASF-2013 were first excluded from the general set of PDBbind v2016. Subsequently, 1,000 complexes were randomly sampled from the refined set as the validation set, following standard practice \cite{stepniewska2018development, zheng2019onionnet}. The remaining general set complexes were then used to construct the training set.
%%==============================================================%%
%%==============================================================%%
\subsection{Bipartite weighted subgraph descriptors}

We encode the geometric and physicochemical properties of each protein--ligand complex by representing the molecular structure as a geometric graph \( G = (V, E) \), where \( V \) is the set of atoms and \( E \) denotes edges representing non-covalent interactions, such as hydrophobic/hydrophilic contacts and intra-/inter-molecular forces. To incorporate domain-specific chemical context, we assign atom-type labels to the vertices using separate coloring schemes for protein and ligand atoms. Protein atoms are labeled with residue-level atom types---such as \(\alpha\)-carbon (CA), \(\beta\)-carbon (CB), and \(\delta_1\)-carbon (CD1)---while ligand atoms are classified using SYBYL atom types, which reflect their chemical environments (e.g., sp\textsuperscript{3} carbon (C.3), aromatic carbon (C.ar), aromatic nitrogen (N.ar), and amide nitrogen (N.am)). This dual atom-type coloring enables a fine-grained description of molecular interactions and structural roles.

Interactions between protein and ligand atoms are explicitly modeled by constructing a bipartite weighted subgraph for each complex, where protein and ligand atoms form two disjoint vertex sets. Let \(\mathcal{T}_p\) and \(\mathcal{T}_l\) denote the sets of atom types in the protein and ligand, respectively. Each atom is represented as a vertex \((\mathbf{r}, t)\), where \(\mathbf{r} \in \mathbb{R}^3\) denotes its spatial coordinate and \(t \in \mathcal{T}_p \cup \mathcal{T}_l\) is its type. 
 The protein and ligand vertex sets are defined as
\[
V_p = \{ (\mathbf{r}_i, t^{(p)}_k) \mid \mathbf{r}_i \in \mathbb{R}^3,\ t^{(p)}_k \in \mathcal{T}_p;\ i = 1, \ldots, N_p \},
\]
\[
V_l = \{ (\mathbf{r}_j, t^{(l)}_{k'}) \mid \mathbf{r}_j \in \mathbb{R}^3,\ t^{(l)}_{k'} \in \mathcal{T}_l;\ j = 1, \ldots, N_l \}.
\]
Edges are introduced between protein and ligand atoms whose Euclidean distance is below a predefined cutoff \(c > 0\). The bipartite edge set is defined as
\[
E_{pl} = \left\{ \Phi\left( \|\mathbf{r}_i - \mathbf{r}_j\|;\ \eta_{kk'} \right) \,\middle|\, (\mathbf{r}_i, t^{(p)}_k) \in V_p,\ (\mathbf{r}_j, t^{(l)}_{k'}) \in V_l,\ \|\mathbf{r}_i - \mathbf{r}_j\| \le c \right\},
\]
where the weight function \(\Phi\) encodes the strength of the interaction. The characteristic interaction distance is defined as
\[
\eta_{kk'} = \eta (r_k + r_{k'}),
\]
where \(r_k\) and \(r_{k'}\) are the van der Waals radii of atom types \(t^{(p)}_k\) and \(t^{(l)}_{k'}\), respectively, and \(\eta > 0\) is a scaling factor. The function \(\Phi(d; \eta_{kk'})\) is chosen to satisfy
\[
\Phi(d; \eta_{kk'}) \to 1 \quad \text{as } d \to 0, \quad \text{and} \quad \Phi(d; \eta_{kk'}) \to 0 \quad \text{as } d \to \infty.
\]
Two common choices for \(\Phi\) are the generalized exponential and Lorentz functions:
\[
\Phi_E(d; \eta_{kk'}) = \exp\left( - \left( \frac{d}{\eta_{kk'}} \right)^\kappa \right), \quad
\Phi_L(d; \eta_{kk'}) = \frac{1}{1 + \left( \frac{d}{\eta_{kk'}} \right)^\kappa}, \quad \kappa > 0.
\]
The resulting bipartite subgraph is formally represented as \(G_{pl} = (V_p \cup V_l,\ E_{pl})\), and serves as a structural representation of the protein-ligand binding interface. Informative geometric features are extracted from the bipartite subgraph by defining a centrality score for each atom that quantifies its interaction density with atoms of a specific paired type. For a given atom \((\mathbf{r}_i, t_k)\) in the protein and a corresponding atom type \(t_{k'}\) in the ligand, we compute the geometric centrality \(C_{kk'}(\mathbf{r}_i)\) as the cumulative interaction strength with all ligand atoms of type \(t_{k'}\) within the cutoff distance. Formally, this is defined as
\[
C_{kk'}(\mathbf{r}_i) = \sum_{(\mathbf{r}_j, t_{k'}) \in V_l} \Phi\left(\|\mathbf{r}_i - \mathbf{r}_j\|;\ \eta_{kk'}\right) \cdot \delta(t_j = t_{k'}),
\]
where \(\delta(t_j = t_{k'})\) is an indicator function equal to 1 if the atom type of \(\mathbf{r}_j\) is \(t_{k'}\), and 0 otherwise. This centrality score captures how strongly a given atom is engaged in interactions with a specific ligand atom type, and is similarly defined for ligand atoms interacting with protein atom types. 
To construct a compact and expressive representation for each bipartite subgraph defined by atom type pair \((t_k, t_{k'})\), we aggregate centrality values computed for all atoms—both protein and ligand—that participate in the subgraph. Let \(C_{kk'} = \{C_{kk'}(\mathbf{r}_i)\}_{i=1}^{n_{kk'}}\) denote the set of centrality values for all relevant atoms associated with the subgraph defined by \((t_k, t_{k'})\). We compute a subgraph-level feature vector by applying a collection of statistical functions:
\[
\mathbf{f}_{kk'} = \mathrm{Stat}(C_{kk'}) \in \mathbb{R}^s,
\]
where \(\mathrm{Stat}(\cdot)\) returns a fixed-length summary vector containing \(s\) statistical descriptors such as summation, mean, median, variance, standard deviation, minimum, and maximum. This unified representation captures the overall structural and interaction characteristics of the subgraph. The resulting tensor of shape \(|\mathcal{T}_p| \times |\mathcal{T}_l| \times s\) serves as a structured and interpretable encoding of the geometric and chemical interactions within the complex.

%%==============================================================%%
%%==============================================================%%

\subsection{Model Architecture}

\begin{figure}[t]
    \centering
    \includegraphics[width=0.98\linewidth]{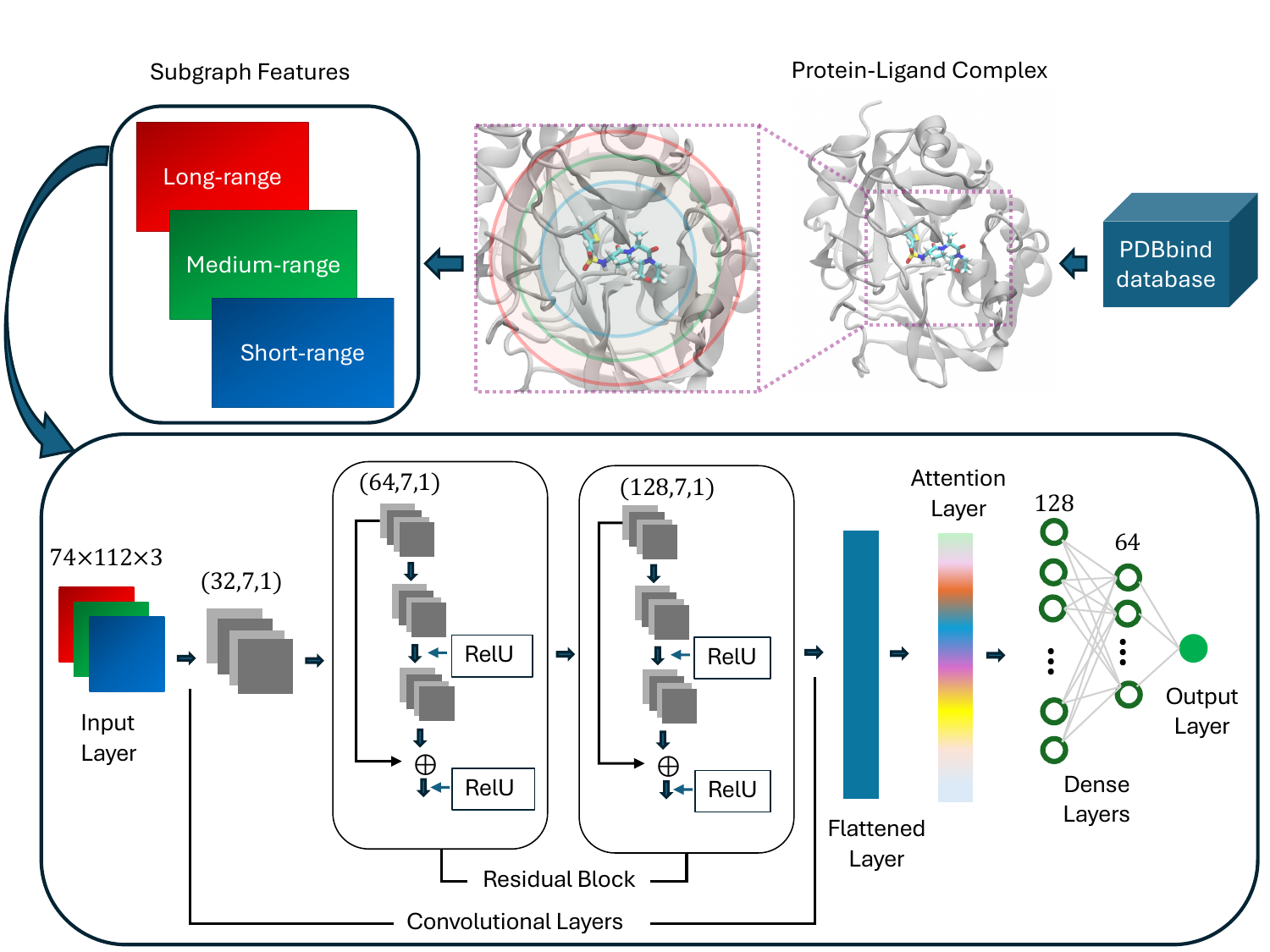}
    \caption{Workflow of the proposed DeepGGL model for protein-ligand binding affinity prediction.}
    \label{fig:deepGGL_workflow}
\end{figure}

The overall workflow for DeepGGL is depicted in Figure~\ref{fig:deepGGL_workflow}. For each protein–ligand complex, geometric graph features are derived from bipartite subgraphs constructed using weighted, atom-type-colored vertices. Atom centrality values are computed for all pairwise protein–ligand atom type combinations and summarized using eight statistical descriptors (count, sum, mean, median, variance, standard deviation, min, max), yielding a feature tensor of size \(37 \times 28 \times 8 = 8{,}288\). Additionally, three distance cutoffs—5~\AA{} (short-range), 10~\AA{} (medium-range), and 15~\AA{} (long-range)—are used to construct three parallel tensors, which are ultimately combined into a final 3D input tensor of shape \(74 \times 112 \times 3\), analogous to an RGB image.

% \subsection{Convolutional Backbone with Residual Blocks}

Prior to being fed into the network, the raw data is standardized using the \texttt{StandardScaler} from the \texttt{scikit-learn} library, which transforms each feature to have zero mean and unit variance. This normalization ensures that all input features contribute equally to the learning process, prevents dominance of features with larger numeric ranges, and improves the stability and convergence speed of gradient-based optimization.

The standardized input tensor is first processed by a 2D convolutional layer with 32 filters, after which it passes through two successive residual blocks.
Residual connections---introduced in the ResNet architecture~\cite{he2016deep}---enable the network to learn a residual mapping 
\(
\mathcal{F}(\mathbf{x}) = \mathcal{H}(\mathbf{x}) - \mathbf{x}
\)
rather than a direct mapping \(\mathcal{H}(\mathbf{x})\). The output of a residual block can thus be expressed as
\[
\mathbf{y} = \mathcal{F}(\mathbf{x}, \{W_i\}) + \mathbf{x},
\]
where \(\mathbf{x}\) is the input to the block, \(\mathbf{y}\) is the output, and \(\{W_i\}\) denotes the learnable parameters. This formulation allows identity mappings to be propagated directly through the network, thereby mitigating vanishing gradient issues, accelerating convergence, and enabling effective training of deeper models. Here, each residual block contains two 2D convolutional layers, with output channels of 64 and 128 in the first and second blocks, respectively. Batch normalization is applied after each convolution to stabilize learning dynamics, and ReLU activation introduces nonlinearity. The skip connections not only preserve gradient flow but also promote feature reuse, allowing subsequent layers to focus on learning complementary refinements rather than redundant low-level representations. This design facilitates deeper and more expressive feature extraction while reducing the risk of network degradation in performance with increasing depth.

% \subsection{Attention Mechanism}

The final convolutional output is flattened and passed through a self-attention mechanism, implemented using the built-in \texttt{Attention} layer in TensorFlow~Keras. This Luong-style dot-product attention mechanism \cite{luong2015effective} dynamically reweights features by computing their pairwise similarities, enabling the model to focus on the most informative spatial patterns within the subgraph-based representation. Formally, for query (\(Q\)), key (\(K\)), and value (\(V\)) matrices derived from the flattened features, the attention output is given by
\[
\mathrm{Attention}(Q, K, V) = \mathrm{softmax}\!\left( \frac{Q K^\top}{\sqrt{d_k}} \right) V,
\]
where \(d_k\) is the dimensionality of the key vectors. This mechanism enhances prediction accuracy by allowing the network to assign higher weights to more relevant feature interactions.

% \subsection{Fully Connected Layers}

Following attention, the model includes two dense layers with 128 and 64 neurons, respectively. Each dense layer is followed by batch normalization and a dropout layer with a rate of 0.1. L2 regularization with a penalty parameter of \(\lambda = 0.01\) is applied to all dense layers to reduce overfitting and improve generalization. The final output layer is a single linear neuron used to regress the predicted binding affinity (\(pK\)).

% \subsection{Loss Function and Optimization}

The model is trained using a custom loss function that combines the Pearson correlation coefficient (\(R_p\)) and the root mean square error (RMSE). The composite loss is defined as:
\[
\mathcal{L} = \alpha(1 - R_p) + (1 - \alpha) \cdot \mathrm{RMSE}, 
\]
which encourages the model to produce predictions that are both accurate in magnitude and highly correlated with the ground truth. 
The parameter \(\alpha \in [0,1]\)
 serves as a weighting factor to balance the contributions of 
\(R_p\) and RMSE.

Stochastic Gradient Descent (SGD) with Nesterov momentum (0.9), a learning rate of 0.001, decay of \(10^{-6}\), and gradient clipping (clipvalue = 0.01) is used as the optimization algorithm. The model is trained with a batch size of 64 for up to 500 epochs. An early stopping mechanism halts training if the validation loss fails to improve by at least 0.001 over 20 consecutive epochs.

The entire model is implemented using the Keras API with TensorFlow as the backend. The only model hyperparameter optimized in this study was the convolutional kernel size and the weighting factor \(\alpha\) in the loss function. To reduce the complexity of hyperparameter tuning and ensure architectural consistency, the same kernel size was applied to all convolutional layers. Kernel size plays a critical role in determining the receptive field of each convolutional filter, thereby controlling the balance between capturing fine-grained local patterns and broader contextual information within the input representation. The optimal kernel size was determined via five-fold cross-validation on the validation set (details provided in the Supporting Information). As shown in Figure~S1, a kernel size of 7 achieved the highest predictive performance, suggesting that this receptive field effectively captures the spatial dependencies relevant to our protein–ligand subgraph features. The optimal value for $\alpha$ was determined by a similar five-fold cross validation. Figure~S2 reveals that $\alpha=0.7$ achieved the best performance. Consequently, a kernel size of 7 and $\alpha=0.7$ was adopted in the final model configuration.

%%==============================================================%%
%%==============================================================%%
\section{Experiments and Results}
\subsection{Evaluation metrics}
To assess the predictive performance of our DeepGGL model, we employed three widely used regression metrics: the Pearson correlation coefficient (\(R_p\)), Root Mean Squared Error (RMSE), and Mean Absolute Error (MAE). These metrics capture different aspects of prediction quality and allow for a comprehensive evaluation of model accuracy and robustness.

% \subsection{Pearson Correlation Coefficient (\(R_p\))}

The Pearson correlation coefficient \(R_p\) measures the linear correlation between the predicted and actual binding affinity values. It is defined as:

\[
R_p = \frac{\sum_{i=1}^{n} (y_i - \bar{y})(\hat{y}_i - \bar{\hat{y}})}{\sqrt{\sum_{i=1}^{n}(y_i - \bar{y})^2} \cdot \sqrt{\sum_{i=1}^{n}(\hat{y}_i - \bar{\hat{y}})^2}},
\]

where \(y_i\) and \(\hat{y}_i\) are the true and predicted binding affinity values, respectively, and \(\bar{y}\) and \(\bar{\hat{y}}\) denote their respective means. \(R_p\) ranges from \(-1\) to \(1\), with \(R_p = 1\) indicating perfect positive correlation, \(R_p = 0\) indicating no correlation, and \(R_p = -1\) indicating perfect negative correlation. A high \(R_p\) value reflects strong predictive consistency with the ground truth.

% \subsection{Root Mean Squared Error (RMSE)}

The RMSE quantifies the average magnitude of prediction error and is particularly sensitive to large deviations. It is given by:

\[
\mathrm{RMSE} = \sqrt{\frac{1}{n} \sum_{i=1}^{n} (y_i - \hat{y}_i)^2}.
\]

Lower RMSE values indicate better predictive accuracy. Because RMSE penalizes larger errors more heavily than smaller ones, it is useful for identifying models that tend to make large deviations from true values.

% \subsection{Mean Absolute Error (MAE)}

The MAE provides a measure of average absolute prediction error, offering a more interpretable and less variance-sensitive alternative to RMSE:

\[
\mathrm{MAE} = \frac{1}{n} \sum_{i=1}^{n} |y_i - \hat{y}_i|.
\]

MAE is less influenced by outliers and provides a straightforward measure of average prediction deviation. Lower MAE values denote better model performance.

% \subsection{Interpretation}

Together, \(R_p\), RMSE, and MAE provide complementary insights: \(R_p\) evaluates the strength of the linear relationship between predictions and ground truth, while RMSE and MAE quantify the magnitude of prediction error. Models with high \(R_p\) and low RMSE/MAE are considered optimal for regression tasks such as binding affinity prediction.

%%==============================================================%%
%%==============================================================%%
% \subsection{Hyperparameter optimization}

%%==============================================================%%
%%==============================================================%%

\subsection{Predictive power}

\begin{figure}[htbp]
  \centering

  % First subfigure
  \begin{subfigure}[b]{0.45\textwidth}
    \centering
    \includegraphics[width=\textwidth]{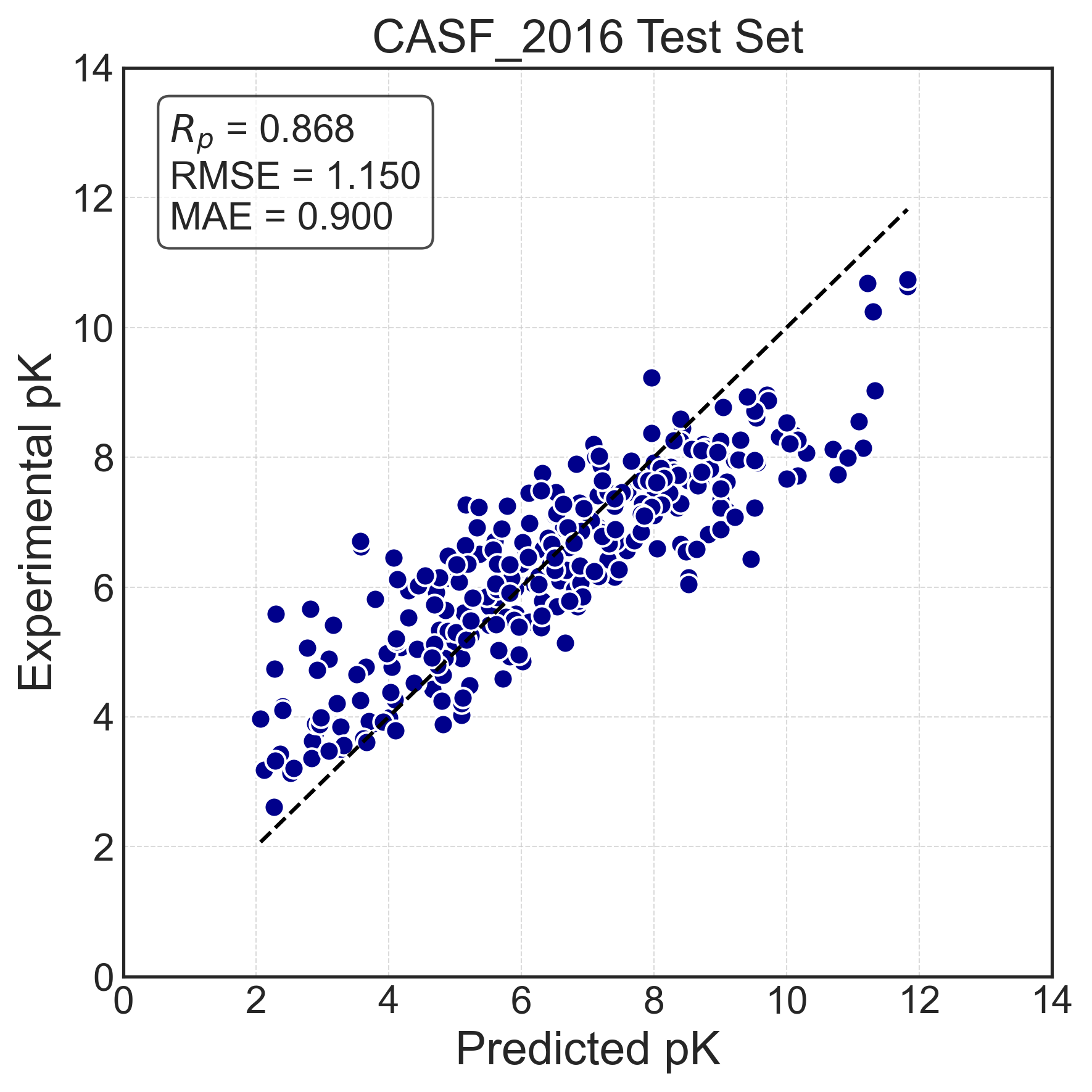}
    % \caption{CASF 2016}
    \label{fig:casf_16_prediction}
  \end{subfigure}
  \hfill
  % Second subfigure
  \begin{subfigure}[b]{0.45\textwidth}
    \centering
    \includegraphics[width=\textwidth]{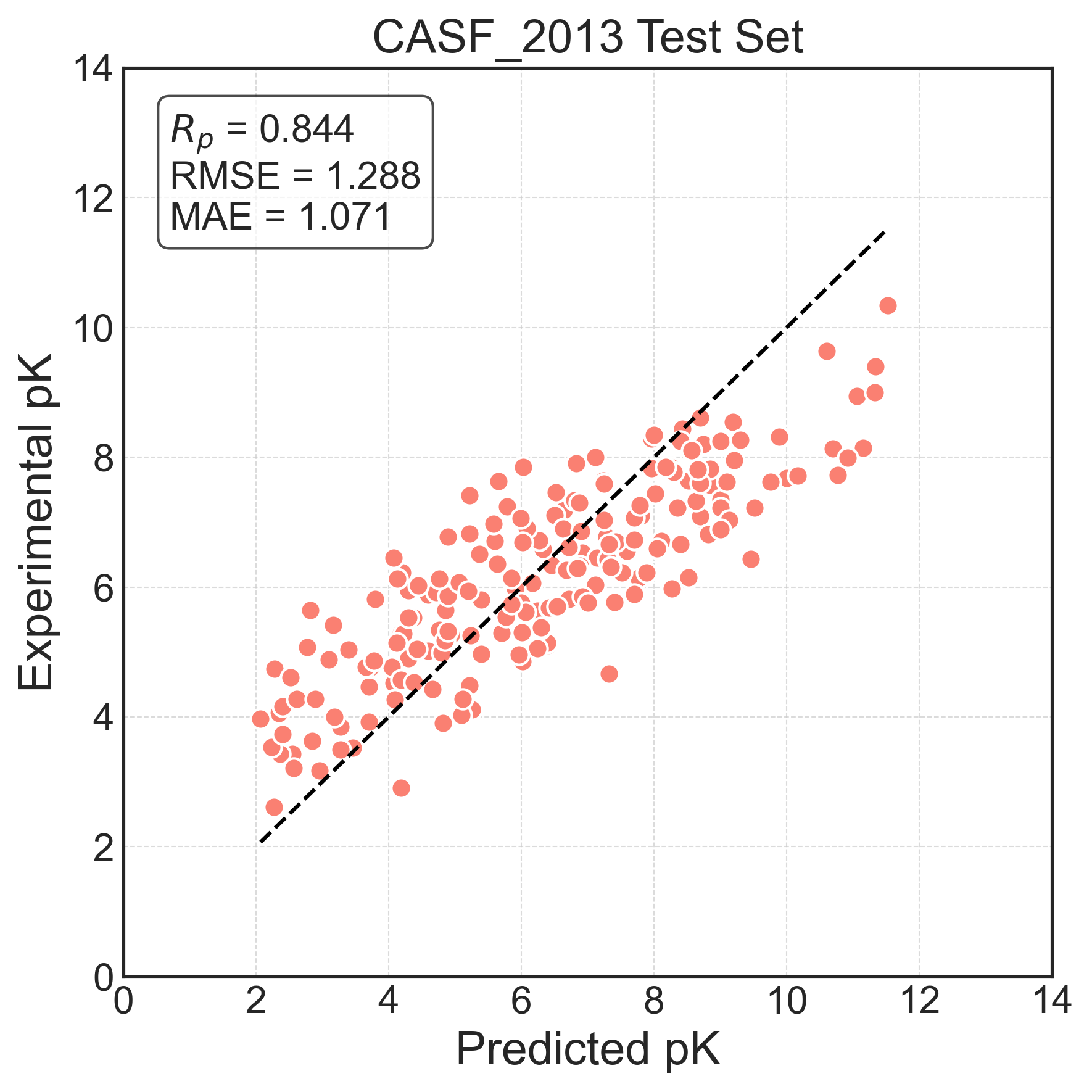}
    % \caption{CASF 2013}
    \label{fig:casf_13_prediction}
  \end{subfigure}

  \caption{Prediction results of DeepGGL on CASF-2016 (left) and CASF-2013 (right) test sets. Each point shows the mean predicted pK for a complex, averaged over five independent training runs, versus its experimentally determined pK.}
  \label{fig:casf_prediction}
\end{figure}

To assess the predictive performance of DeepGGL, we trained the model independently five times and averaged the resulting predictions to obtain the final output for each protein–ligand complex. This ensemble approach reduces variance and improves robustness. Figure \ref{fig:casf_prediction} presents scatter plots of the mean predicted values versus experimentally measured binding affinities on the CASF-2016 and CASF-2013 benchmark test sets. DeepGGL demonstrates strong predictive performance, achieving Pearson correlation coefficients exceeding 0.80 across all datasets. Specifically, the model yields an $R_p$ of 0.810 and an RMSE of 1.158 on the validation set; an $R_p$ of 0.868 and an RMSE of 1.149 on CASF-2016; and an $R_p$ of 0.844 with an RMSE of 1.288 on CASF-2013 (see Table \ref{tab:model_performance}). These results highlight DeepGGL's ability to generalize across datasets while maintaining both high correlation and low prediction error.

\begin{table}[htbp]
\begin{threeparttable}
 \centering
		\caption{Results of DeepGGL on different datasets.\Bstrut}
		%\footnotesize
		\begin{tabular*}{\columnwidth}{@{\hspace{1em}\extracolsep\fill}llll@{\hspace{1em}\extracolsep\fill}}
			%\toprule
			%\begin{tabular}{l c c}
			%\begin{tabular}{c d{4.2} d{2.2} d{2.2} d{2.2} d{2.2}}
			\hline
			Dataset & $R_p$  & RMSE & MAE \Tstrut\Bstrut\\
			\hline
                training set & 0.996 & 0.212 & 0.165 \Tstrut\Bstrut\\
                validation set & 0.810 & 1.158 & 0.882  \Tstrut\Bstrut\\
                CASF-2016 core set & 0.868 &1.150 & 0.900 \Tstrut\Bstrut\\
                CASF-2013 core set & 0.844 &1.288 & 1.071\Tstrut\Bstrut\\
			\hline
		\end{tabular*}
		\label{tab:model_performance}
    \begin{tablenotes}
		\item Note: The error metrics are in pKd units.
	\end{tablenotes}
 \end{threeparttable}
\end{table}

\subsubsection{Performance on CASF-2016 Benchmark}

The CASF-2016 core set is a widely used benchmark for evaluating the generalization capability of scoring functions across diverse protein–ligand complexes. As summarized in Table \ref{tab:performance_comp_CASF_2016}, DeepGGL achieves the best performance among both traditional machine learning (ML) and deep learning (DL) methods, with an $R_p$ of 0.868, RMSE of 1.50, and MAE of 0.90.

Compared to recent graph-based models such as EGNA ($R_p = 0.842$) and BAPA ($R_p = 0.819$), DeepGGL shows a notable improvement in both correlation and error metrics. It also outperforms 3D spatially-aware convolutional models like OnionNet ($R_p = 0.816$) and OnionNet-2 ($R_p = 0.864$), which rely on distance-based grids but lack subgraph-level geometric representation. Sequence-based models such as DeepDTAF and CAPLA show weaker performance on this structure-based benchmark, underscoring the limitations of omitting 3D spatial information.

Traditional ML-based methods like RF-Score v3 and AGL-Score yield lower correlation coefficients ($R_p = 0.812$ and $R_p = 0.833$, respectively) and higher RMSE values, reflecting the advantage of end-to-end learning in DL architectures such as DeepGGL. These results affirm that incorporating geometric and topological information from colored subgraphs into a CNN-attention framework enhances the model's capacity to learn discriminative patterns for binding affinity prediction.

\begin{table}[htbp]
\begin{threeparttable}
	% \begin{center}
 \centering
		\caption{Performance comparison of DeepGGL and state-of-the-art SFs on CASF-2016 benchmark.\Bstrut}
		%\footnotesize
		\begin{tabular*}{\columnwidth}{@{\hspace{1em}\extracolsep\fill}lllll@{\hspace{1em}\extracolsep\fill}}
			%\toprule
			%\begin{tabular}{l c c}
			%\begin{tabular}{c d{4.2} d{2.2} d{2.2} d{2.2} d{2.2}}
			\hline
			Type &Method & $R_p \uparrow$ & RMSE $\downarrow$ & MAE $\downarrow$\Tstrut\Bstrut\\
			\hline
   		\multirow[t]{4}{*}{ML}	
                &PerSpect \cite{meng2021persistent} &0.840 &1.265 & - \Tstrut\Bstrut\\
                &AGL-Score \cite{nguyen2019agl} &0.833 &1.271 & - \Tstrut\Bstrut\\
                &RF-Score v3 \cite{li2015improving} & 0.812 & 1.395 &1.121\Tstrut\Bstrut\\
                &PLEC-Linear \cite{wojcikowski2019development} & 0.760 & 1.454 &1.138\Tstrut\Bstrut\\
            \multirow[t]{9}{*}{DL}	 
            &DeepGGL & \textbf{0.868} & \textbf{1.150} & \textbf{0.900} \Tstrut\Bstrut\\
            &OnionNet-2 \cite{wang2021onionnet} & 0.864 & 1.164 & - \Tstrut\Bstrut\\
            &EGNA \cite{xia2023leveraging} & 0.842 & 1.258 & 0.980 \Tstrut\Bstrut\\
            &K$_{\mathrm{DEEP}}$ \cite{jimenez2018k}& 0.820 & 1.270 & -\Tstrut\Bstrut\\
            &BAPA \cite{seo2021binding} & 0.819 & 1.308 & 1.021 \Tstrut\Bstrut\\
			&OnionNet \cite{zheng2019onionnet}& 0.816 & 1.278 & 0.984 \Tstrut\Bstrut\\
            &Pafnaucy \cite{stepniewska2018development} & 0.780 & 1.420 &1.130\Tstrut\Bstrut\\
            &CAPLA$^\text{*}$ \cite{jin2023capla}&0.843 &1.200 &0.966 \Tstrut\Bstrut\\
            &DeepDTAF$^\text{*}$ \cite{wang2021deepdtaf}&0.789 &1.355 &1.073 \Tstrut\Bstrut\\
			\hline
		\end{tabular*}
		\label{tab:performance_comp_CASF_2016}
	% \end{center}
	\begin{tablenotes}
		\item Note: The error metrics are in pKd units, and the value in bold indicates the best outcome for the respective metric. The methods marked by an asterisk indicate a 1D sequence-based model. 
	\end{tablenotes}
 \end{threeparttable}
\end{table}

\subsubsection{Performance on CASF-2013 Benchmark}

On the CASF-2013 core set, DeepGGL also achieves the best performance among all evaluated methods, with an $R_p$ of 0.844, RMSE of 1.288, and MAE of 1.071 (Table \ref{tab:performance_comp_CASF_2013}). This benchmark, although older, still poses significant challenges due to its diverse composition and lower sequence similarity to training sets.

DeepGGL outperforms OnionNet-2 ($R_p = 0.821$) and OnionNet ($R_p = 0.782$), further supporting the utility of learning features from atom-level subgraph statistics over simple spatial representations. Additionally, graph-based models such as CAPLA ($R_p = 0.770$) and Pafnaucy ($R_p = 0.700$) perform worse, suggesting that explicit modeling of protein–ligand subgraph interactions at multiple distance scales, as done in DeepGGL, contributes to higher generalization capability.

The performance gap between DeepGGL and sequence-based methods such as DeepDTAF (with a relatively low $R_p = 0.608$) is especially pronounced in CASF-2013, indicating that sequence-only models may fail to capture the nuanced spatial and physicochemical interactions necessary for accurate affinity prediction in diverse protein–ligand complexes.

Together, these evaluations demonstrate that DeepGGL not only achieves state-of-the-art performance on benchmark datasets but also generalizes effectively across structurally and compositionally diverse protein–ligand complexes. The integration of subgraph-based geometric features with a deep residual CNN and attention mechanism provides a powerful framework for binding affinity prediction.

\begin{table}[htbp]
\begin{threeparttable}
	% \begin{center}
 \centering
		\caption{Performance comparison of DeepGGL and state-of-the-art SFs on CASF-2013 benchmark.\Bstrut}
		%\footnotesize
		\begin{tabular*}{\columnwidth}{@{\hspace{1em}\extracolsep\fill}lllll@{\hspace{1em}\extracolsep\fill}}
			%\toprule
			%\begin{tabular}{l c c}
			%\begin{tabular}{c d{4.2} d{2.2} d{2.2} d{2.2} d{2.2}}
			\hline
			Type &Method & $R_p \uparrow$ & RMSE $\downarrow$ & MAE $\downarrow$\Tstrut\Bstrut\\
			\hline
            \multirow[t]{2}{*}{ML}
            &PerSpect &0.793 &1.435 & - \Tstrut\Bstrut\\
            &AGL-Score &0.792 &1.447 & - \Tstrut\Bstrut\\
            \multirow[t]{6}{*}{DL}
   			&DeepGGL & \textbf{0.844} & \textbf{1.288} & \textbf{1.071} \Tstrut\Bstrut\\
            &OnionNet-2 & 0.821 & 1.357 & - \Tstrut\Bstrut\\
                
			&OnionNet & 0.782 & 1.503 & 1.208 \Tstrut\Bstrut\\
                &Pafnaucy & 0.700 & 1.620 &1.320\Tstrut\Bstrut\\
                &CAPLA$^\text{*}$ & 0.770 & 1.446 &1.154\Tstrut\Bstrut\\
                &DeepDTAF$^\text{*}$ &0.608 &2.103 &1.737\Tstrut\Bstrut\\
			\hline
		\end{tabular*}
		\label{tab:performance_comp_CASF_2013}
	% \end{center}
	\begin{tablenotes}
		\item Note: The error metrics are in pKd units, and the value in bold indicates the best outcome for the respective metric. The methods marked by an asterisk indicate a 1D sequence-based model. 
	\end{tablenotes}
 \end{threeparttable}
\end{table}

%%==============================================================%%
%%==============================================================%%
\subsection{Multi-range interaction}
Protein–ligand binding is governed by a complex interplay of atomic interactions that occur over various spatial ranges. To effectively capture this diversity, DeepGGL integrates geometric features extracted at three distinct distance cutoffs: short-range (5\AA), medium-range (10\AA), and long-range (15\AA). This multi-range feature representation enables the model to encode both local atomic contacts and more distal structural interactions that collectively influence binding affinity.

\begin{figure}
    \centering
    \includegraphics[width=0.98\linewidth]{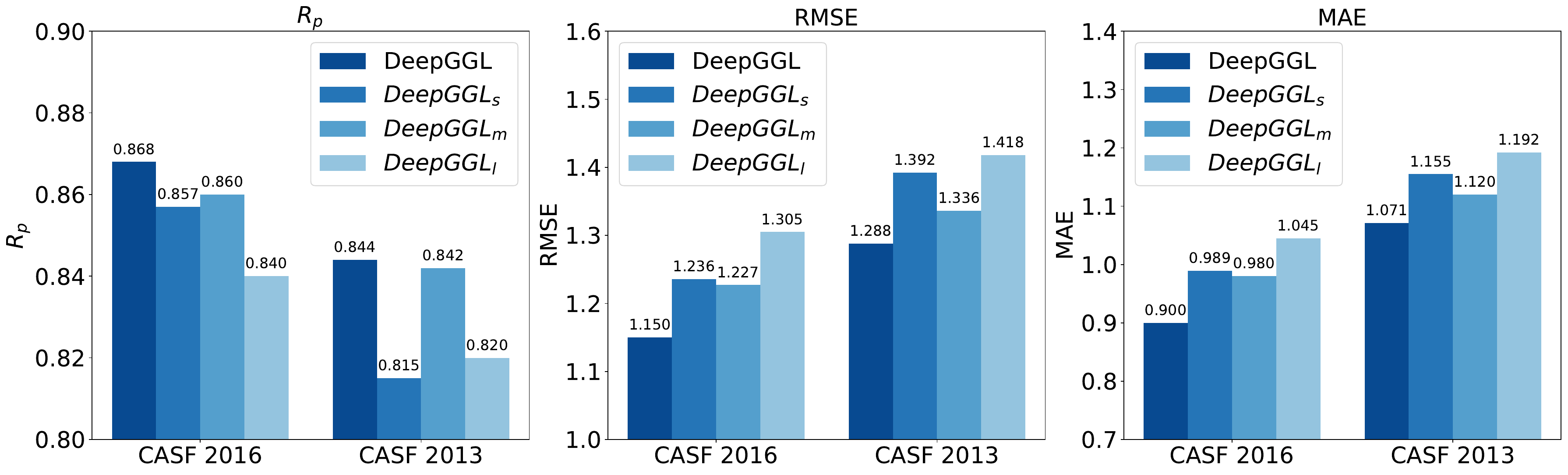}
    \caption{Comparison of DeepGGL, DeepGGL$_s$, DeepGGL$_m$, and DeepGGL$_l$ on CASF 2016 and CASF 2013 datasets.}
    \label{fig:cutoff_comparison}
\end{figure}

To systematically evaluate the contribution of each spatial scale, we constructed three ablation models— DeepGGL$_\text{s}$, DeepGGL$_\text{m}$, and DeepGGL$_\text{l}$—each trained with features derived from only one of the distance cutoffs. All inputs were reshaped to a uniform dimension of $(74, 112, 1)$ to ensure consistent model architecture across variants. Figure~\ref{fig:cutoff_comparison} illustrates the comparative performance of these models alongside the full DeepGGL model, which integrates features across all three ranges.

On the CASF-2016 benchmark, DeepGGL$_\text{m}$ (medium-range features) achieves the best single-scale performance with an $R_p$ of 0.8600, RMSE of 1.2265, and MAE of 0.9795. This suggests that interactions within a 10,\AA{} radius are particularly informative for predicting binding affinity. DeepGGL$_\text{s}$ (short-range, 5,\AA) performs comparably well, with an $R_p$ of 0.8571, though it shows slightly higher RMSE (1.2362) and MAE (0.9889), indicating that short-range contacts alone are somewhat less predictive. DeepGGL$_\text{l}$ (long-range, 15,\AA) yields a lower $R_p$ of 0.8400 and higher RMSE (1.3054), suggesting that while long-range interactions contribute valuable context, they are less discriminative when used in isolation.

A similar trend is observed on the CASF-2013 benchmark. DeepGGL$_\text{m}$ again performs best among the ablation models ($R_p = 0.8423$, RMSE = 1.3356, MAE = 1.1203), followed closely by DeepGGL$_\text{s}$ ($R_p = 0.8146$, RMSE = 1.3919, MAE = 1.1549). DeepGGL$_\text{l}$ trails with an $R_p$ of 0.8201 and the highest RMSE and MAE values (1.4177 and 1.1918, respectively).

Importantly, the full DeepGGL model—incorporating all three spatial scales—achieves the best overall performance on both benchmarks (see Table \ref{tab:model_performance}), underscoring the synergistic value of integrating multi-range features. These findings highlight the complementary roles of short-, medium-, and long-range interactions in capturing the structural determinants of protein–ligand binding. The multi-range representation equips DeepGGL with a more holistic view of the complex interaction landscape, leading to improved generalization and predictive accuracy.

%%==============================================================%%
%%==============================================================%%
\subsection{Generalization ability}

To evaluate the generalization capability of DeepGGL beyond the CASF benchmarks, we conducted independent testing on two external datasets: the CSAR NRC-HiQ set and the PDBbind v2019 `hold-out set'. These datasets represent protein–ligand complexes not included during model training or validation and span diverse structural and temporal characteristics. Assessing model performance on these sets offers a robust indication of its ability to generalize to previously unseen and more recent protein–ligand pairs, which is crucial for real-world applications in drug discovery.

The Community Structure-Activity Resource (CSAR) dataset \cite{dunbar2011csar} is a high-quality benchmark curated to evaluate scoring functions for protein–ligand docking and affinity prediction. We specifically utilize the CSAR NRC-HiQ subset, which contains 343 experimentally resolved protein–ligand complexes. This subset was originally divided into two groups based on deposition time in the Protein Data Bank (PDB), consisting of 176 and 167 complexes, respectively. Following the protocol of Xia et al.\cite{xia2023leveraging}, we merged the two subsets and removed any complexes that overlapped with the PDBbind v2016 training set to avoid data leakage. The resulting filtered set contains 49 unique complexes, providing a stringent test of model generalization to structurally diverse and temporally disjoint samples.

As shown in Table \ref{tab:CSAR-HiQ_results}, DeepGGL achieves the highest Pearson correlation coefficient ($R_p = 0.764$) and the lowest RMSE (1.512) among all evaluated models, with a competitive MAE (1.211). It outperforms state-of-the-art methods such as EGNA ($R_p = 0.750$), PLEC-Linear ($R_p = 0.741$), and K$_{\mathrm{DEEP}}$ ($R_p = 0.710$), indicating that its learned geometric representations are highly transferable. Notably, although EGNA achieves a slightly lower MAE (1.190), its lower correlation and higher RMSE suggest reduced consistency across the dataset. Traditional machine learning models such as RF-Score v3 perform markedly worse ($R_p = 0.532$, RMSE = 1.917), demonstrating the superiority of deep learning-based feature extraction in generalization tasks.

\begin{table}[htbp]
\begin{threeparttable}
	% \begin{center}
 \centering
		\caption{Performance of DeepGGL on CSAR-NRC-HiQ Dataset.\Bstrut}
		%\footnotesize
		\begin{tabular*}{\columnwidth}{@{\hspace{1em}\extracolsep\fill}llll@{\hspace{1em}\extracolsep\fill}}
			%\toprule
			%\begin{tabular}{l c c}
			%\begin{tabular}{c d{4.2} d{2.2} d{2.2} d{2.2} d{2.2}}
			\hline
			Method & $R_p \uparrow$ & RMSE $\downarrow$ & MAE $\downarrow$\Tstrut\Bstrut\\
			\hline
   			DeepGGL & \textbf{0.764} & \textbf{1.512} & 1.211 \Tstrut\Bstrut\\
                EGNA & 0.750 & 1.536 & \textbf{1.190} \Tstrut\Bstrut\\
                PLEC-Linear & 0.741 & 1.656 &1.242\Tstrut\Bstrut\\
                K$_{\mathrm{DEEP}}$ & 0.710 & 1.630 & 1.257\Tstrut\Bstrut\\
                BAPA & 0.669 & 1.700 & 1.346 \Tstrut\Bstrut\\
			OnionNet & 0.661 & 1.714 & 1.319 \Tstrut\Bstrut\\
                Pafnaucy & 0.620 & 1.903 &1.602\Tstrut\Bstrut\\
                RF-Score v3 & 0.532 & 1.917 &1.433\Tstrut\Bstrut\\
			\hline
		\end{tabular*}
		\label{tab:CSAR-HiQ_results}
	% \end{center}
	\begin{tablenotes}
		\item Note: The error metrics are in pKd units, and the value in bold indicates the best outcome for the respective metric.
             The performances of other models are adopted from \cite{xia2023leveraging}. 
	\end{tablenotes}
 \end{threeparttable}
\end{table}

To further assess the performance of DeepGGL on newer structural data, we evaluated it on the PDBbind v2019 hold-out set, as used in prior studies \cite{jones2021improved}. This set comprises 222 protein–ligand complexes from the PDBbind v2019 refined set that were deposited after the PDBbind v2016 release and are therefore disjoint from the CASF-2016 benchmark. This dataset provides a realistic simulation of predicting binding affinity for novel and unseen protein–ligand structures, making it highly relevant for prospective applications.

DeepGGL again demonstrates strong generalization, outperforming other models in all three evaluation metrics on this dataset (Table \ref{tab:v2019_hold-out_results}). It achieves the highest $R_p$ of 0.557, along with the lowest RMSE (1.267) and MAE (1.001). Compared to DeepFusion ($R_p = 0.545$) and Pafnucy ($R_p = 0.528$), DeepGGL exhibits better predictive power and lower prediction errors. The performance gap is even more pronounced against older models such as K$_{\mathrm{DEEP}}$ ($R_p = 0.487$, RMSE = 1.424), confirming that DeepGGL’s multi-range geometric encoding enables better generalization to evolving protein–ligand structures.

Across both independent test sets, DeepGGL consistently achieves superior performance, reflecting its ability to generalize well beyond the training distribution. The results validate that DeepGGL’s design—leveraging geometric and algebraic graph-based features across multiple interaction ranges—is robust to structural novelty and dataset drift, a critical advantage for practical deployment in computational drug discovery pipelines.

\begin{table}[htbp]
\begin{threeparttable}
	% \begin{center}
 \centering
		\caption{Performance of DeepGGL on PDBbind v2019 Hold-Out set.\Bstrut}
		%\footnotesize
		\begin{tabular*}{\columnwidth}{@{\hspace{1em}\extracolsep\fill}llll@{\hspace{1em}\extracolsep\fill}}
			%\toprule
			%\begin{tabular}{l c c}
			%\begin{tabular}{c d{4.2} d{2.2} d{2.2} d{2.2} d{2.2}}
			\hline
			Method & $R_p \uparrow$ & RMSE $\downarrow$ & MAE $\downarrow$\Tstrut\Bstrut\\
			\hline
                DeepGGL & \textbf{0.557} & \textbf{1.267} & \textbf{1.001} \Tstrut\Bstrut\\
                DeepFusion & 0.545 & 1.338 & 1.074  \Tstrut\Bstrut\\
                Pafnucy & 0.528 & 1.381 & 1.106 \Tstrut\Bstrut\\
                K$_{\mathrm{DEEP}}$ & 0.487 & 1.424 & 1.135 \Tstrut\Bstrut\\
			\hline
		\end{tabular*}
		\label{tab:v2019_hold-out_results}
	% \end{center}
	\begin{tablenotes}
		\item Note: The error metrics are in pKd units, and the value in bold indicates the best outcome for the respective metric.
	\end{tablenotes}
 \end{threeparttable}
\end{table}

\subsection{Model ablation study}

To better understand the contribution of key architectural components in DeepGGL, we conducted an ablation study by systematically removing the residual block and attention mechanism (architectures are in the supporting information). The analysis was performed across four benchmark datasets: CASF-2016, CASF-2013, CSAR NRC-HiQ, and the PDBbind v2019 hold-out set. Table~S1 summarizes the performance in terms of the Pearson correlation coefficient ($R$), reflecting the quality of predicted binding affinities relative to experimental values.

{\it Effect of Residual Connection}:
Removing the residual connection from DeepGGL leads to a consistent but modest drop in performance across all datasets. Specifically, the correlation on CASF-2016 decreased from 0.868 to 0.857, and on the PDBbind v2019 hold-out set from 0.557 to 0.547. These results highlight the residual block's role in facilitating deeper network architectures by enabling better gradient flow and mitigating the vanishing gradient problem. Its inclusion appears particularly important for datasets like PDBbind v2019, which contain more structurally diverse and recent complexes, suggesting that residual connections help the model generalize to complex unseen data.

{\it Effect of Attention Mechanism}:
To assess the contribution of the attention mechanism, we removed it while retaining all other components. Interestingly, DeepGGL without attention slightly outperformed the full model on CASF-2013 (0.845 vs. 0.844) and CSAR NRC-HiQ (0.769 vs. 0.764), albeit by a narrow margin. However, on CASF-2016 and PDBbind v2019 hold-out, performance decreased slightly, especially in the latter case (0.544 vs. 0.557). These observations indicate that the attention mechanism is most beneficial in scenarios where complex spatial patterns and long-range dependencies are critical, such as newer datasets with more diverse structures. Its effect may be dataset-dependent, with diminishing returns in benchmark sets that exhibit more uniform interactions.

{\it Effect of Removing Both Components}:
When both the residual block and the attention mechanism are removed, the model shows a clear degradation in performance across all benchmarks. The Pearson correlation drops to 0.861 on CASF-2016 and to 0.507 on the PDBbind v2019 hold-out set—an appreciable loss of predictive power compared to the full model. This decline confirms that these components work synergistically to enhance the model’s capacity to capture non-linear and hierarchical interactions across spatial scales.

Overall, both the residual connection and attention mechanism play important roles in enhancing DeepGGL's performance and generalization ability. The residual block contributes to stable and deep learning, especially in structurally diverse scenarios, while the attention mechanism provides adaptability in capturing long-range and context-dependent geometric features. Their combined effect yields the highest performance across the majority of benchmarks, affirming their design relevance in the DeepGGL architecture.

\section{Conclusion}
In this study, we introduced DeepGGL, a deep convolutional neural network designed to predict protein–ligand binding affinity with high accuracy. By incorporating bipartite weighted subgraph descriptors at multiple interaction ranges, DeepGGL effectively captures both geometric and chemical features of protein–ligand complexes. Comprehensive benchmarking on CASF-2013 and CASF-2016 datasets demonstrates that DeepGGL outperforms existing state-of-the-art methods. Further validation on independent datasets, including CSAR NRC-HiQ and the PDBbind v2019 hold-out set, confirms the model's strong generalization capabilities. The robustness of DeepGGL to structural diversity and dataset shift underscores its practical applicability in real-world drug discovery settings. Ablation studies further reveal that architectural components such as residual connections and attention mechanisms are instrumental in enhancing model performance. Overall, DeepGGL represents a significant step forward in structure-based affinity prediction, offering a reliable and scalable approach for accelerating computational drug design.

\section*{Acknowledgements}
This work is supported in part by funds from the National Science Foundation (NSF: \# 2516126, \# 2151802, and \# 2534947), Kennesaw State University startup fund, and the Kennesaw State University high-performance computing cluster.

\section*{Competing interests}
The authors declare no competing interests.

\section*{Supplementary information}

The supplementary material contains the hyperparameter optimization and additional results on independent test sets and ablation models. 

\bibliographystyle{unsrt}
\bibliography{refs}

\newpage
\section*{Supporting Information}

\setcounter{section}{0}

\setcounter{table}{0}
\renewcommand{\thetable}{S\arabic{table}}
\setcounter{figure}{0}
\renewcommand\thefigure{S\arabic{figure}}

\section{Model hyper-parameters}
\begin{figure}[htbp]
    \centering
    \includegraphics[width=0.6\linewidth]{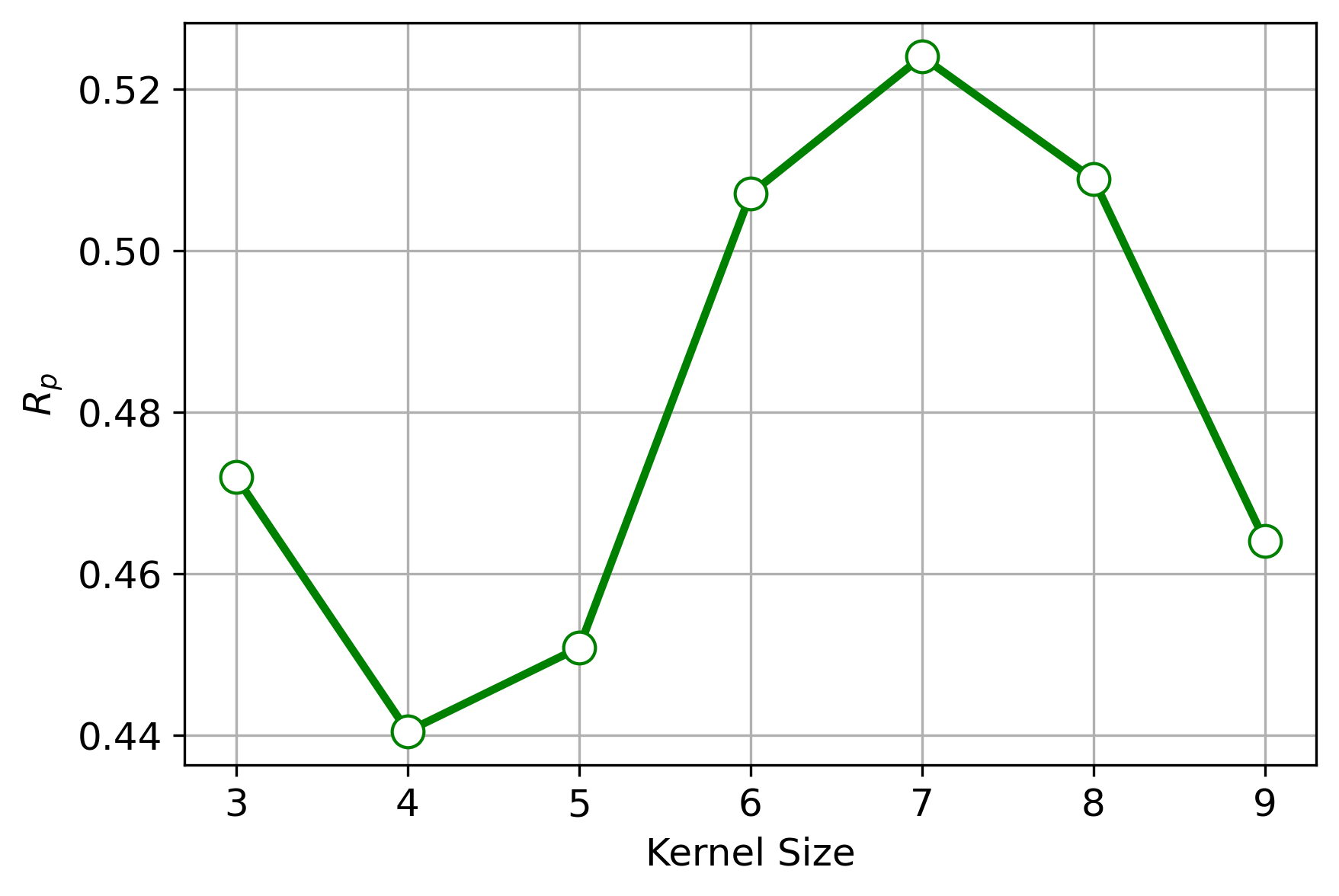}
    \caption{5-fold cross-validation results of DeepGGL in Pearson correlation ($R_p$) for varying kernel size. The kernel size of 7 shows the best performance.}
    \label{fig:kernel_size_cv}
\end{figure}

\begin{figure}[htbp]
    \centering
    \includegraphics[width=0.6\linewidth]{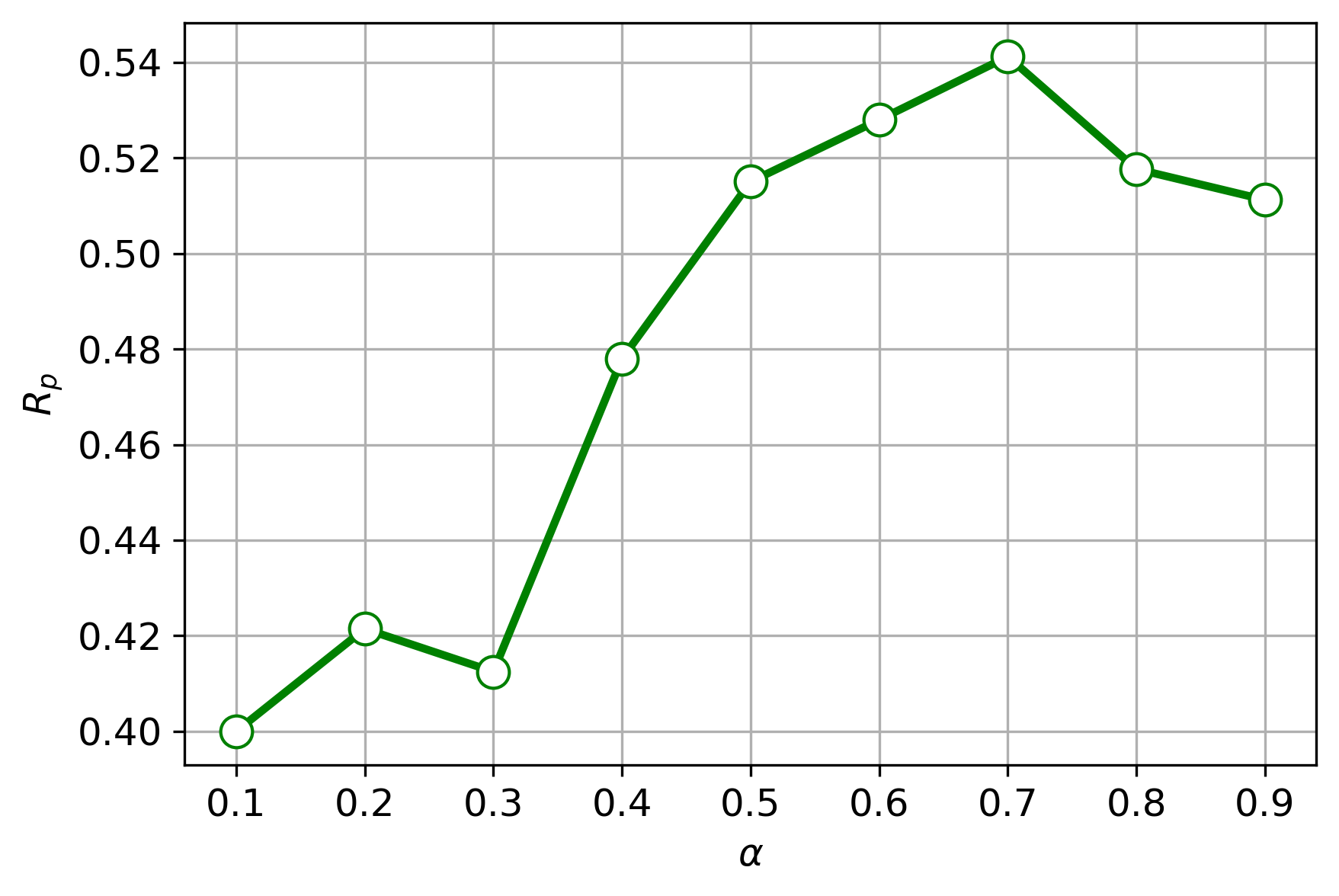}
    \caption{5-fold cross-validation results of DeepGGL in Pearson correlation ($R_p$) for varying loss function weight parameter $\alpha$. The value $\alpha=0.7$ shows the best performance.}
    \label{fig:alpha_cv}
\end{figure}

To identify the optimal kernel size and loss function weight parameter ($\alpha$), we performed a systematic hyperparameter search using five-fold cross-validation (CV) on the validation set, while keeping all other model parameters fixed as specified in the main text. Specifically, we employed the RepeatedKFold strategy from scikit-learn with five repeats, and repeated the entire CV procedure three times, reporting the average performance across runs. For kernel size optimization, we fixed $\alpha = 0.5$ and varied the kernel size from 3 to 9. As shown in Figure \ref{fig:kernel_size_cv}, a kernel size of 7 achieved the highest Pearson correlation coefficient. Using this optimal kernel size, we next optimized $\alpha$ by varying its value between 0.1 and 0.9 in increments of 0.1. As illustrated in Figure \ref{fig:alpha_cv}, the best performance was obtained with $\alpha = 0.7$.

\section{Results on independent test sets}
\begin{figure}[htbp]
  \centering

  % First subfigure
  \begin{subfigure}[]{0.45\textwidth}
    \centering
    \includegraphics[width=\textwidth]{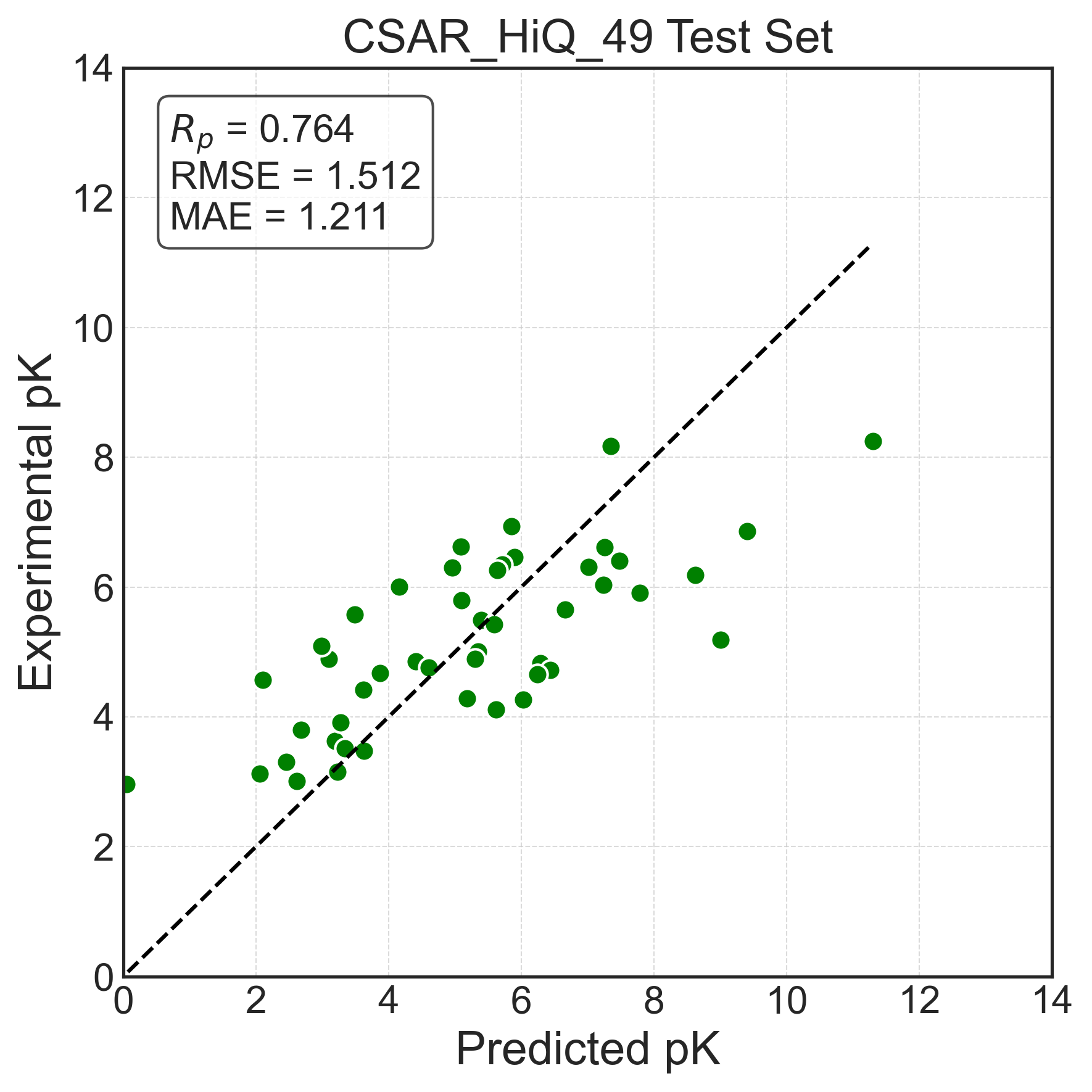}
    % \caption{CASF 2016}
    \label{fig:csar_prediction}
  \end{subfigure}
  \hfill
  % Second subfigure
  \begin{subfigure}[]{0.45\textwidth}
    \centering
    \includegraphics[width=\textwidth]{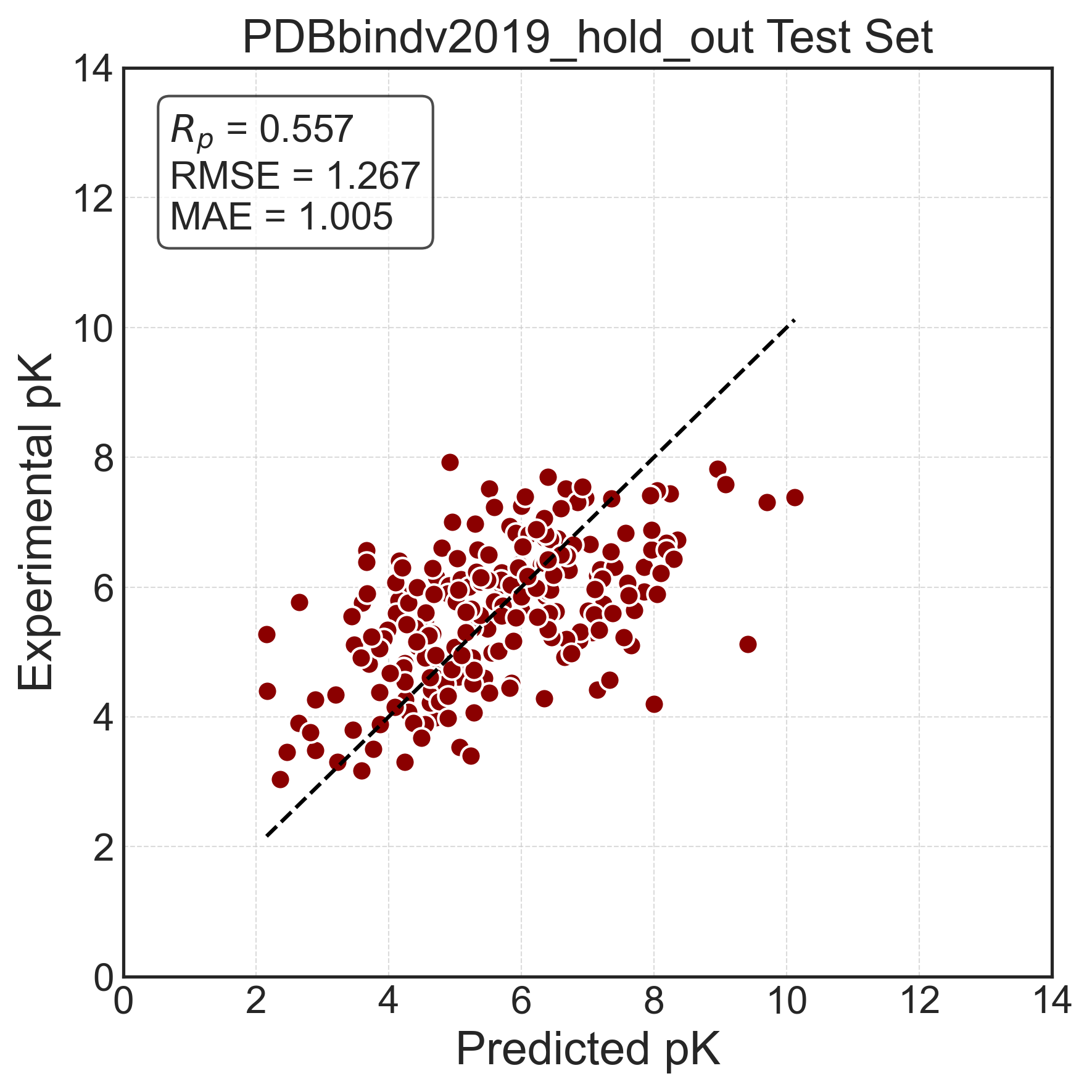}
    % \caption{CASF 2013}
    \label{fig:v19_prediction}
  \end{subfigure}

  \caption{Prediction results of DeepGGL on CSAR-NRC-HiQ (left) and PDBbind v2019 holdout (right) test sets. Each point shows the mean predicted pK for a complex, averaged over five independent training runs, versus its experimentally determined pK.}
  \label{fig:additional_test_prediction}
\end{figure}

Scatter plots of the mean predicted values of five independent runs versus experimentally measured binding affinities on the two external datasets, the CSAR NRC-HiQ set and the PDBbind v2019 `hold-out set', are presented in Figure \ref{fig:additional_test_prediction}.

\section{Ablation model architecture and results}
\begin{figure}[t]
    \centering
    \includegraphics[width=0.98\linewidth]{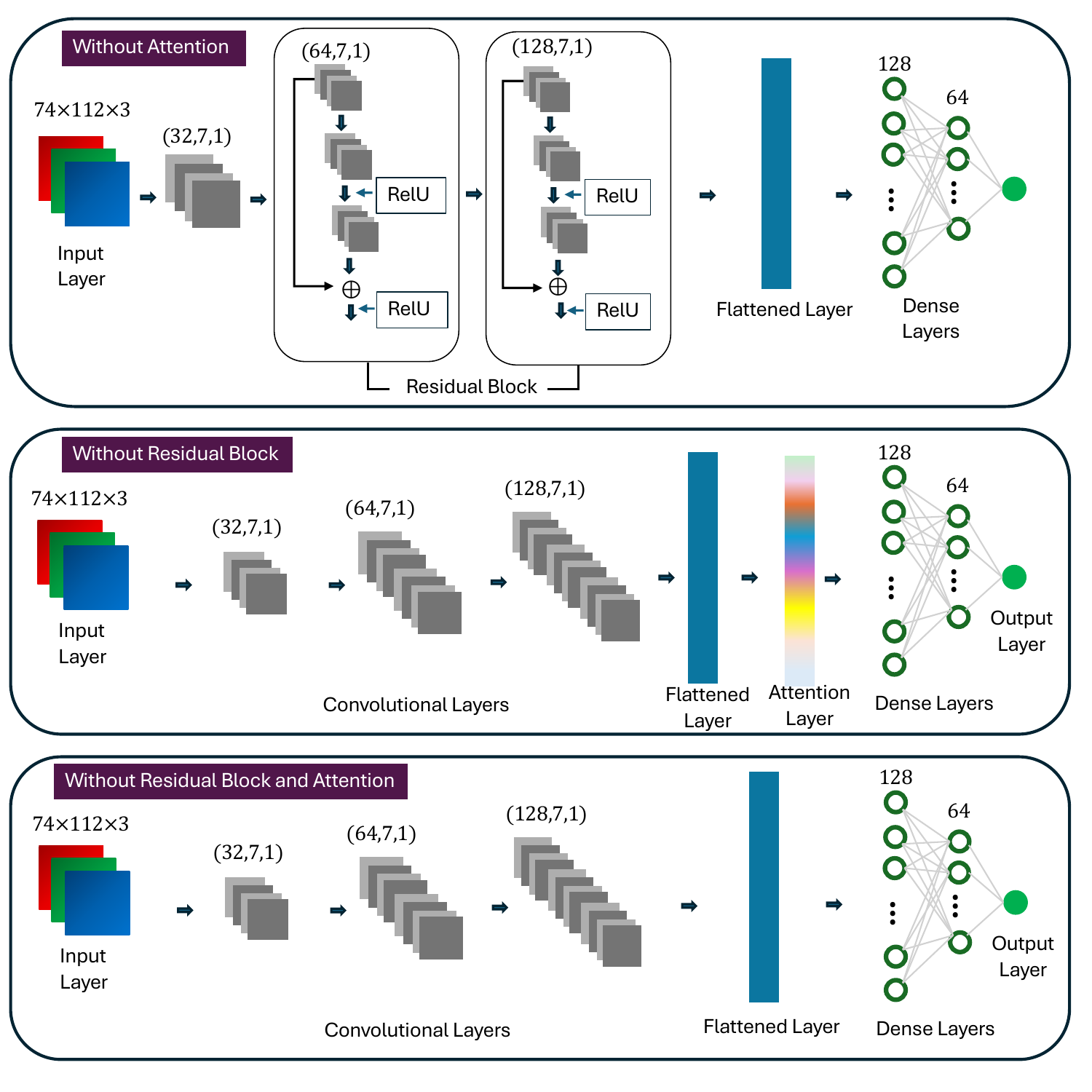}
    \caption{DeepGGL ablation model architectures.}
    \label{fig:ablation_model_architecture}
\end{figure}

Architectures of various ablation models are provided in Figure \ref{fig:ablation_model_architecture} and their performance on four benchmark test sets is listed in Table \ref{tab:ablation_r_only}.

\begin{table*}[hbpt]
\begin{center}
\caption{Model ablation study: Pearson correlation coefficient ($R$) across four benchmark datasets.\Bstrut}
\tabcolsep=10pt
\begin{tabular*}{\columnwidth}{@{\hspace{1em}\extracolsep\fill}lcccc@{\hspace{1em}\extracolsep\fill}}
\hline
Model Variant & CASF 2016 & CASF 2013 & CSAR-HiQ & v2019-holdout \Tstrut\Bstrut\\
\hline
DeepGGL & \textbf{0.868} & 0.844 & 0.764 & \textbf{0.557} \Tstrut\\
Without residual block & 0.857 & 0.835 & 0.767 & 0.547 \\
Without attention mechanism & 0.863 & \textbf{0.845} & \textbf{0.769} & 0.544 \\
Without attention and residual block & 0.861 & 0.840 & 0.756 & 0.507 \Bstrut\\
\hline
\end{tabular*}
\label{tab:ablation_r_only}
\end{center}
\end{table*}

\end{document}